\documentstyle[aps,multicol,epsfig]{revtex}
\begin{document}
\draft
\title{Depletion driven adsorption of colloidal rods onto a hard wall}

\author{Richard P.~Sear$^{\dag}$}

\address{FOM Institute for Atomic and Molecular Physics, 
Kruislaan 407, 1098 SJ Amsterdam, The Netherlands}

\address{$^{\dag}$Address from October 1997:
Department of Chemistry and Biochemistry, The University of
California, Los Angeles, 405 Hilgard Avenue,
Los Angeles, California 90095-1569}

\date{\today}

\maketitle

\vspace{0.1in}

\begin{abstract}
In a mixed suspension of rods and small polymer coils, the rods adsorb
onto a hard wall in contact with the suspension.
This adsorption is studied in the low density of rods limit.
It is driven by depletion forces and is much stronger for long rods than for
spheres. This is shown by means of exact, numerical, calculations and
an approximate theory.
\end{abstract}

\pacs{67.70.+n,~82.70.Dd,~61.25.Hq}

\begin{multicols}{2}
\narrowtext

\section{Introduction}

Most of the properties of a colloidal suspension depend on the
shape of its constituent particles.
Certainly, the bulk
phase behaviour of colloidal particles is sensitive to their shape,
as well as to the strength of any attractive interactions.
Spheres behave differently from
rods which in turn behave differently from discs.
This is true in the presence \cite{schoot92,sear97} and in the
absence \cite{frenkel91,allen94} of attractive interactions.
However, a description of the bulk phase behaviour does not complete
the description of a suspension. We might also like to know, for example,
its behaviour near the walls of its container. We study this behaviour
here for rodlike colloidal particles in the presence of nonadsorbing
polymer, and compare this behaviour to that of spherical particles.
A pure suspension of rods does not adsorb onto a hard wall, in the
absence of significant van der Waals attractions;
here we will assume that they are negligible.
Indeed, the
density of rods near the wall is below the bulk density. However, the
rods in a mixed suspension of rods and small nonadsorbing (onto either
the rods or the wall) polymer coils do adsorb onto the wall. This
adsorption is driven by the increase in the volume available to the
polymer coils when a rod is near a wall, the so-called depletion
forces \cite{gast83,lekkerkerker92,russel89,poon95,lekkerkerker95}.
We find much stronger adsorption for rods than has been found
for spheres \cite{kaplan94b,sober95,dinsmore96}.

The depletion driven adsorption of spherical colloidal particles onto
a wall has been extensively
studied \cite{kaplan94b,sober95,dinsmore96,kaplan94,ohshima97}.
These studies were primarily experimental but comparison was made
with theories for depletion forces \cite{asakura54,vrij76} and
qualitative agreement found.
Both the sphere and the wall exclude the centres of mass of the polymer
coils from a volume which extends up to about the radius of gyration
of the polymer away from the surface of the sphere or wall. This is true
so long as the radius of gyration is not too much larger than the radius
of the sphere \cite{eisenriegler96}. When a sphere approaches the wall,
the volume the sphere excludes to the polymer and the volume the wall
excludes to the polymer overlap. Thus the total volume denied to the
polymer coils goes down and so their translational entropy goes up.
For polymer coils rather smaller than the spheres,
free energies of adsorption of
up to $6k_BT$ have been obtained \cite{kaplan94b,sober95}.
$k_B$ is Boltzmann's constant and $T$ is the temperature.

Rodlike colloids are not uncommon, examples are the tobacco mosaic
virus \cite{fraden95} and synthetic colloidal rods
\cite{lekkerkerker95,buining94}. The effect of adding polymer to their
suspensions has been studied, at least in the bulk
\cite{lekkerkerker95,fraden95,buitenhuis95}.
As far as we are aware,
no experiments comparable to those for spheres near a wall have
been performed.
However, Buitenhuis {\it et al} \cite{buitenhuis95} report
a thin layer of the nematic phase of the rods against the wall of their
capillary. This is at least suggestive of some attraction
between the rods and the capillary.
The adsorption of rods does not
seem to have been much studied theoretically. An exception is
the work of Matsuyama {\it et al} \cite{matsuyama97}, who studied
a model of rods with discrete orientations.
However, the adsorption of polymers which have
some degree of rigidity has been studied, see Ref. \cite{linden96} and
references therein. The results are consistent with those found here,
as the rigidity of the polymer is increased, so does the adsorption.

In the following section we first derive the exact statistical mechanical
expression for the density profile of rods near a wall, as a function
of the polymer density. We use this expression in two ways: The first
is by evaluating it exactly using Monte Carlo integration and the second
is by deriving a simple analytical approximation to it. Example
profiles of the density of rods near the wall,
are shown and discussed in Section III. We end with a conclusion,
Section IV.

\section{Theory}

\begin{figure}
\begin{center}
\epsfig{file=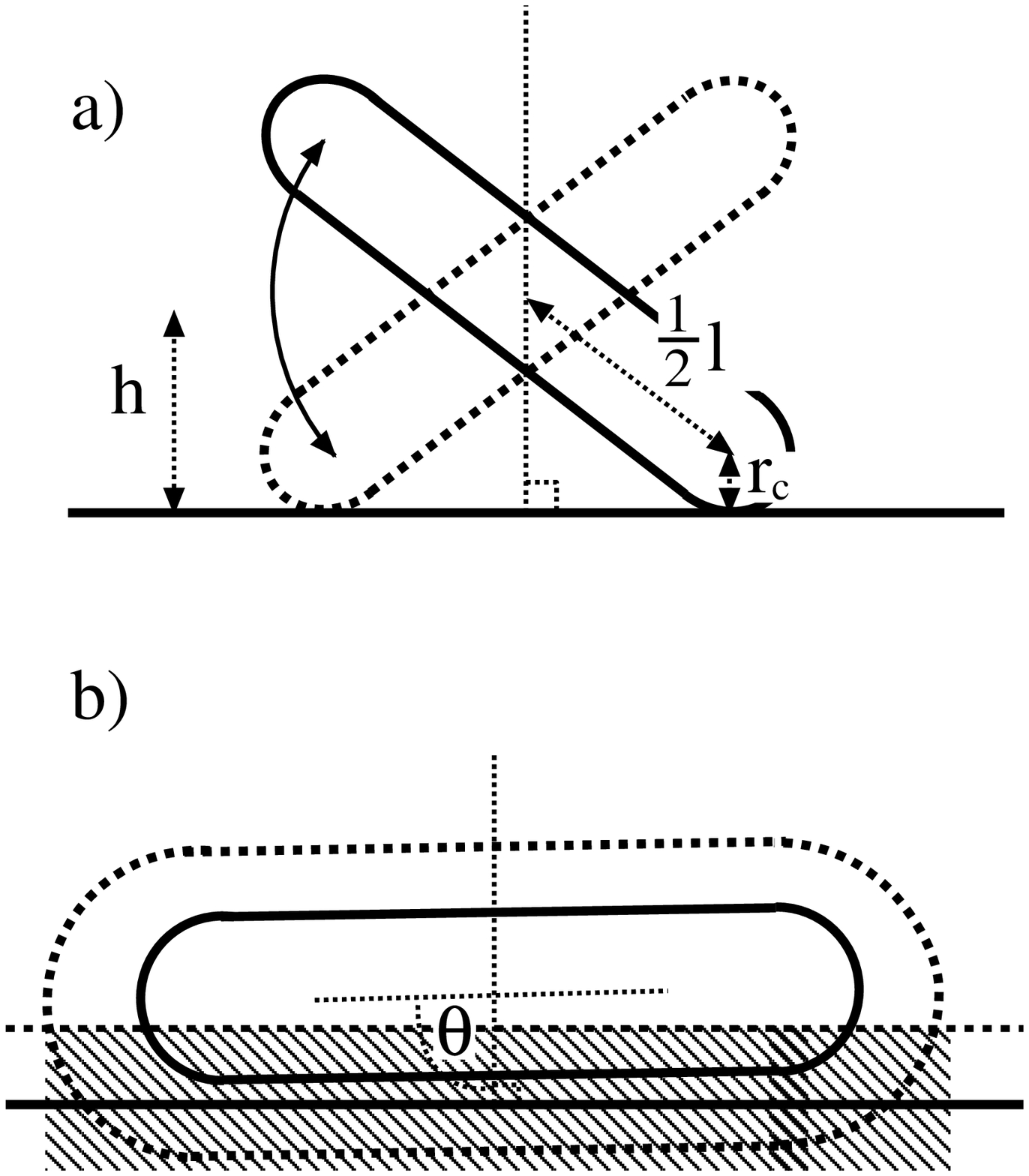, width=3.0in}
\end{center}
\vspace*{0.6in}
\caption{
Schematics of a hard spherocylinder near a hard wall.
(a) The solid spherocylinder shows it at the smallest angle possible
without overlap of the spherocylinder with the wall
and the dashed spherocylinder shows it the largest angle.
The solid arc with arrows denotes the range of angles
available to the spherocylinder at that height.
(b) The cross-section of the overlap of the volume a rod excludes
to the polymer with the volume the wall excludes to the polymer is
shown as the shaded area.
The solid lines are the surface of the spherocylinder and the wall, and
the dashed lines are the surfaces of the volumes they exclude to the
polymer.
}
\label{fig1}
\end{figure}

We start by defining our models for the rod, the polymer coil and the wall.
The interactions between polymer coils are neglected, i.e., a fluid
of just polymer coils is simply an ideal gas. This leaves us with just
three interactions in order to describe our system: The rod--wall,
polymer--wall and rod--polymer interactions.
The rod--wall interaction is that of a hard spherocylinder with a
smooth hard wall. The spherocylinder is of length $l$ and diameter
$d=2r_c$.
Thus, no point on
the centreline of the rod, a line of length $l$ which runs along the
centre of the cylindrical portion of the spherocylinder,
may be within $r_c$ of the wall, see Fig. \ref{fig1}(a).
Note that, as usual \cite{vroege92}, the length $l$ is the length of
the cylindrical part of the spherocylinder; its total length is $l+d$.
For the purposes of the polymer--wall and polymer--rod interactions,
a polymer coil is considered to be a hard sphere of radius $r_p$;
this is the so-called Asakura--Oosawa model
\cite{asakura54,vrij76,warren94,lekkerkerker94}.
Then the position of a polymer coil is entirely determined by the position 
of its centre of mass. This centre of mass is prevented
from coming within $r_p$ of the wall and from $r_p+r_c$ from any point
on the centreline of the rod.
The rod excludes polymer molecules from a spherocylindrical volume
of length $l$ and diameter $2(r_c+r_p)$.

Now that we have defined our model we write down the 
partition function $\Xi$ for a system of one colloidal rod
and a fluid of polymer molecules at an activity $z_p$,
\begin{eqnarray}
\Xi&=&\sum_{N_p}\frac{z_p^{N_p}}{N_p!}\int
{\rm d}{\bf r}{\rm d}\omega{\rm d}{\bf r}^{N_p}
\exp[-\beta u_{rw}({\bf r},\omega)]\times\nonumber\\
&&\prod_{i=1,N_p}\exp[-\beta u_{pw}({\bf r}_i)
-\beta u_{rp}({\bf r},\omega,{\bf r}_i)],
\label{xi}
\end{eqnarray}
where the coordinates of the centre of mass of the rod are denoted by
${\bf r}$, and its
orientation by $\omega$.
The coordinates of all $N_p$ polymer coils are denoted by
${\bf r}^{N_p}$ and the coordinates of the $i$th polymer coil
by ${\bf r}_i$. As usual $\beta$ is related to the temperature $T$
by $\beta=1/k_BT$, where $k_B$ is Boltzmann's constant.
The polymer's activity $z_p$ is related to its chemical
potential $\mu_p$ by $z_p=\Lambda^{-1}\exp(\beta\mu_p)$, where
$\Lambda$ is the thermal volume of a polymer coil.
$u_{rw}$, $u_{pw}$ and
$u_{rp}$ are the energies of interaction of
the rod with the wall, a polymer molecule with the wall, and the
rod with a polymer molecule, respectively.
They are all hard-core interactions and so are zero unless the
particles overlap, in which case they are infinite.
As there are no polymer--polymer interactions, Eq. (\ref{xi}) simplifies to
\begin{eqnarray}
\Xi&=&
\int {\rm d}{\bf r}{\rm d}\omega
\exp\left[-\beta u_{rw}({\bf r},\omega)\right]
\times \nonumber \\
&&\sum_{N_p}\frac{z_p^{N_p}}{N_p!}
\left(\int {\rm d}{\bf r}_1
\exp\left[
-\beta u_{pw}({\bf r}_1)
-\beta u_{rp}({\bf r},\omega,{\bf r}_1)\right]
\right)^{N_p}.\nonumber\\
\label{xi2}
\end{eqnarray}

The integrand of the part of Eq. (\ref{xi2}) which is within
parentheses is 0 if the polymer sphere is within $r_p$ of either the wall
or the surface of the spherocylinder and 1 otherwise. Therefore, the
integral within parentheses is equal to the volume of the system which is
farther than $r_p$ from both the wall and the rod's surface, the free
volume for the polymer $V_f({\bf r},\omega)$.
If in addition we recognise the sum of Eq. (\ref{xi2}) as
being just the expansion of an exponential function, we have
\begin{equation}
\Xi=
\int {\rm d}{\bf r}{\rm d}\omega
\exp\left[-\beta u_{rw}({\bf r},\omega)+ z_p V_f({\bf r},\omega)
\right].
\label{xi3}
\end{equation}
The volume $V_f$ may be written as a sum of three parts,
\begin{equation}
V_f({\bf r},\omega)=V_b-v_{exc}+v_o({\bf r},\omega),
\label{vf}
\end{equation}
where $V_b$
is the total volume available to a polymer in the absence of the rod and
$v_{exc}=\pi lr_e^2 + (4/3)\pi r_e^3$, the volume excluded by
the rod to the polymer molecules; $r_e=r_c+r_p$, the radius of the
excluded volume spherocylinder around a rod.
$v_o$ is the volume of overlap
of the volume excluded to the polymer by the rod and the volume excluded
to the polymer by the wall, see Fig. \ref{fig1}(b).
Thus if the rod is far from the wall then the regions excluded to the polymer
by the rod and by the wall do not overlap and $v_o=0$ but if
the rod is close to the wall then these two excluded volumes overlap and
so $v_o>0$. The total volume available to the polymer has increased.
It is this increase in the free volume, or to put it another way, the
reduction in the volume excluded to the polymer, which is the driving
force for adsorption of the rod onto the wall. The adsorption is not driven
by a direct rod--wall interaction as is usually the case.
Using Eq. (\ref{vf}) we can rewrite Eq. (\ref{xi3}) as
\begin{eqnarray}
\Xi&=&
\exp\left[z_p(V_b-v_{exc})\right]\times\nonumber\\
&&\int {\rm d}{\bf r}d\omega
\exp\left[-\beta u_{rw}({\bf r},\omega)
+z_p v_o({\bf r},\omega)\right].
\label{xi4}
\end{eqnarray}
As the wall is smooth both $u_{rw}$ and $v_o$ are functions
only of the angle between the surface normal and the axis of the rod,
$\theta$, and of
the distance between the centre of mass of the rod and the nearest
point of the wall, $h$.

Now that we have integrated over the polymer coordinates, $\Xi$ is a
one particle partition function and so the
probability of finding the rod at coordinates
$(h,\theta)$ is proportional to 
the integrand of Eq. (\ref{xi3}). We now consider an ideal gas of
rods, i.e., rods at a density so low that interactions between the rods
have a negligible effect. Then the rods are independent of each other
and the density of rods is again
proportional to the integrand of Eq. (\ref{xi3}).
If the bulk density of the ideal gas of rods is $\rho_b$ then the density
of rods $\rho(h,\theta)$ is given by
\begin{equation}
\rho(h,\theta)=\rho_b\exp\left[-\beta u_{rw}(h,\theta)
+z_p v_o(h,\theta)\right].
\label{rho}
\end{equation}
This is exact in the low density of rods limit and is the equation we will
use to determine how strongly the rods adsorb onto the surface, in the
presence of the polymer. We are interested in the $h$ not the angle
dependence of the density of rods and so we integrate Eq. (\ref{rho})
over $\theta$
\begin{equation}
\rho(h)=\rho_b\frac{1}{2}
\int_{-1}^{1} {\rm d}(\cos\theta)
\exp\left[-\beta u_{rw}({\bf r},\theta)
+z_p v_o({\bf r},\theta)\right],
\label{rhoh}
\end{equation}
where the factor of a half is the normalisation of the integration
over $\theta$.
Note that Eq. (\ref{rhoh}) has exactly the same
form for rods near a wall with a direct rod--wall attractive interaction;
the term $-z_pk_BTv_o({\bf r},\theta)$
is an effective rod--wall attractive interaction.
It is attractive because if the rod is far from the wall $v_0=0$ and then
as the rod nears the wall $v_0$ becomes positive, thus tending
to increase the density of rods.
Equation (\ref{rhoh}) is directly
related to the change in free energy $\Delta F(h)$ 
when a rod is brought from within the bulk to a height $h$
\begin{equation}
\Delta F =-k_BT\ln\left(\rho(h)/\rho_b\right).
\label{freeh}
\end{equation}
This quantity has actually been measured experimentally for spheres
\cite{kaplan94b,sober95}. As all the interactions are athermal
the energy is zero and the free energy $\Delta F$ is simply an entropy
times $T$.
Finally, from Eq. (\ref{rhoh}) the adsorption $\Gamma$ is easily calculated
using \cite{rowlinson82}
\begin{equation}
\Gamma=\int_0^\infty {\rm d}h
(\rho(h)-\rho_b).
\label{gamma}
\end{equation}

\subsection{Rod near a wall}

It is instructive to consider just a hard spherocylinder near a hard wall;
the $z_p=0$ limit of Eq. (\ref{rhoh}). The potential $u_{rw}$ is zero
as long as the rod and wall do not overlap. For this to be true the
lowest point of the centreline must be at least $r_c$ above the wall, as the
surface of the spherocylinder is $r_c$ away from the centreline. The
lowest point of the centreline of the spherocylinder is always one
of its two ends, see Fig. \ref{fig1}(a). Therefore, $u_{rw}$ is zero
providing both ends of the centreline are at least $r_c$ above the wall
and is infinity otherwise. Then, for $z_p=0$, we can perform
the integration over $\theta$ of Eq. (\ref{rhoh}), to obtain the
density of spherocylinders as a function of height
\begin{equation}
\rho(h)=\left\{
\begin{array}{ll}
\rho_b2(h-r_c)/l & (h-r_c)\le l/2\\
&\\
\rho_b & (h-r_c)>l/2.
\end{array}\right. 
\label{rhoz0}
\end{equation}
The density of centres
of mass of the spherocylinders decreases linearly
as the wall is approached due
to the restricted orientations of a rod close to a wall.

\subsection{Numerical evaluation of $v_o$}

The overlapping excluded volume $v_o(h,\theta)$ is the volume of overlap of a
spherocylinder of length $l$ and radius $r_e$ with a half-space.
We evaluate it numerically using a Monte Carlo integration
scheme \cite{press92}. Briefly, first a randomly located point within
the spherocylinder is generated, in a coordinate frame fixed on the
spherocylinder.
Generating a point within a spherocylinder is done by first selecting
either the cylindrical part or the endcaps.
The probability of selecting the cylinder (endcaps) is given by the
fraction of the total volume of the spherocylinder which is part
of the cylinder (endcaps).
In the cylindrical part, the point's coordinate along the axis is
just a random number between $-l/2$ and $l/2$ and its
coordinates perpendicular to the axis are those of a random point
within a disc. In the endcaps the point is obtained by
first generating a random point within a sphere.
Then if the coordinate of this point along the axis of the spherocylinder
is positive $l/2$ is added to it and if it is negative $l/2$ is subtracted.
Generating a point randomly within a sphere or disc
is done with a routine of Appendix G of Ref. \cite{allen87}.
Now we have a randomly chosen point within the spherocylinder, in a
frame fixed on the spherocylinder. The spherocylinder is then rotated to
an angle of $\theta$ with the normal of the wall.
If the point's height above the wall is then less than $r_p$ then it is
within that part of the volume excluded by the spherocylinder which
is also excluded by the wall and hence is part of $v_o$. If it is above
$r_p$ it is not. If many such points are generated then the fraction of
points which lie below $r_p$ is an approximation to the
fractional overlap $v_o(h,\theta)/v_{exc}$.

\subsection{Approximate theory}

Here we derive a simple analytical approximation for $\rho(h)$.
In the presence of polymer the highest density of rods will be near to the
wall, as then $v_o$ is largest. Therefore, an approximation should be accurate
in this region. The volume $v_o$ is quite small, a fraction of $r_e^3$,
if the angle $\theta$ is small, i.e, the rod is nearly perpendicular
to the wall. However, $v_o$ is large, a fraction of $lr_e^2$,
if the rod is nearly parallel to the wall and close to it, $h<r_e+r_p$. This
is assuming that $l\gg r_e$. In fact as $v_o$ appears as the
argument of the exponential function in Eq. (\ref{rhoh}) for the density,
the density of rods close and nearly parallel to the wall increases
as $\exp(l)$. As we shall see in the next section this can lead to very
strong adsorption.

\begin{figure}
\begin{center}
\epsfig{file=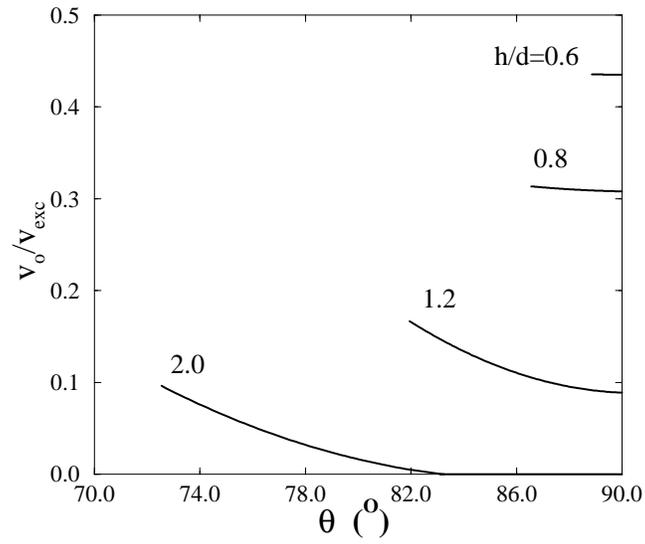, width=3.7in}
\end{center}
\vspace*{-0.7in}
\caption{
The fraction of the volume excluded by the rod which overlaps with
that of the wall, as a function of $\theta$ (in degrees)
for several values of $h/d$.
Each curve is labelled by the value of $h/d$.
For all curves, $l/d=10$ and $r_p=r_c$. 
The curves are plotted over the range in $\theta$ for which the hard core of
the spherocylinder does not overlap with the wall.
}
\label{fig2}
\end{figure}

So, we require an approximation accurate for rods close to, and hence
necessarily almost parallel to the wall.
In Fig. \ref{fig2} we have plotted $v_o/v_{exc}$ as a function of $\theta$
for a number of heights $h$. For small $h$ the spherocylinder
is restricted to a small range of angles by the hard-core
spherocylinder--wall interaction. In addition $v_o$ is not a strongly varying
function of $\theta$, especially near the wall.
Therefore, we approximate $v_o$ for the whole
range of accessible values of $\theta$ by its value for $\theta=90^{\circ}$.
Then, our approximate expression for $\rho(h)$ is
\begin{equation}
\rho(h)=\left\{
\begin{array}{ll}
\rho_b\frac{2(h-r_c)}{l}\exp\left[z_pv_o(h,90^{\circ})\right]
& (h-r_c) \le l/2\\
&\\
\rho_b
&  (h-r_c) > l/2.
\end{array}\right.
\label{rhoa}
\end{equation}
$v_o(h,90^{\circ}$)
is easy to calculate. It consists of two parts, the volume
of the cylinder of radius $r_e$ at a height $h$ which is below
a height of $r_p$ and the volume of a sphere of the same radius which is below
this height. So,
\begin{equation}
v_o(h,90^{\circ})=\left\{
\begin{array}{ll}
l\left[r_e^2
\cos^{-1}\left(\frac{h-r_p}{r_e}\right)\right.&\\
\left. -(h-r_p)\left( r_e^2-(h-r_p)^2 \right)^{1/2} \right]
&\\
+\frac{\pi}{3}\left[2r_e^3-3r_e^2(h-r_p)\right.&\\
+\left. (h-r_p)^3\right]&h\le r_e+r_p\\
&\\
0& h> r_e+r_p.
\end{array}\right.
\label{vo90}
\end{equation}
Equation (\ref{vo90}) is restricted to $r_c\ge r_p$ and so requires
generalisation if $r_p>r_c$. However, the Asakura--Oosawa model
\cite{poon95,asakura54,vrij76} of
the rod--polymer interaction is best for small polymer coils, with
a radius of gyration no larger than that of the colloidal
particle \cite{vrij76,eisenriegler96,sear97b}. 
Therefore, we will not perform
calculations with $r_p>r_c$ and so do not require a more general
expression.

For $h>r_e+r_p$, $v_o(h,90^{\circ})=0$ and the approximation of Eq.
(\ref{rhoa}) yields the same value for $\rho(h)$ as in the absence of
polymer, Eq. (\ref{rhoz0}). However, for small and large (near $\pi$)
$\theta$'s $v_o$ is still non-zero (see Fig. \ref{fig2})
and so Eq. (\ref{rhoa}) is an underestimate.
As $h$ approaches $r_c$ the arc over which the rod may rotate tends
to zero and then the approximation of Eq. (\ref{rhoa}) becomes
increasingly accurate.
For small $r_p$ the rod must be quite close for the excluded
volume of a near parallel rod to overlap with that of the wall.
So, reducing $r_p$ will improve the accuracy
of Eq. (\ref{rhoa}).
Also, if $l$ is large then the arc over which the rod can rotate decreases and
so our approximation becomes more accurate.
In the opposite limit: $l\rightarrow0$, the approximation of Eq. (\ref{rhoa})
is exact; as it is in the $z_p=0$ limit.

\section{Results and discussion}

\begin{figure}
\begin{center}
\epsfig{file=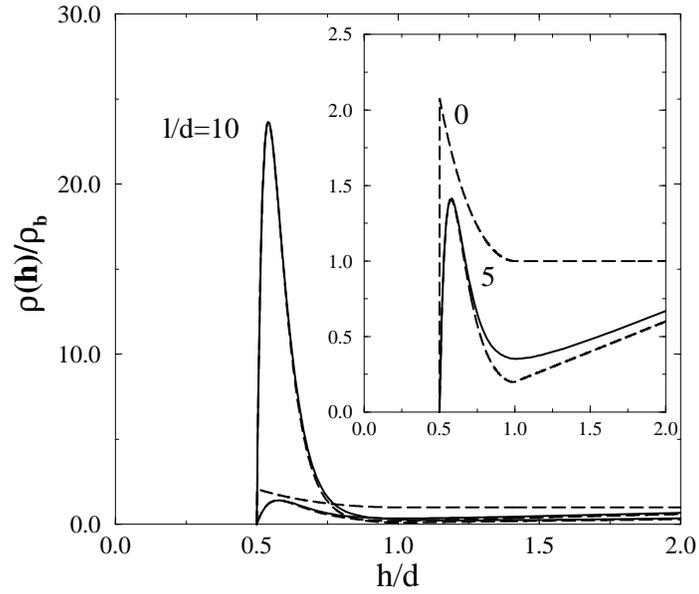, width=3.7in}
\end{center}
\vspace*{-0.7in}
\caption{
The ratio of the density of rods $\rho(h)$ to their bulk density $\rho_b$,
as a function of height $h$ above the wall.
The 2 solid curves are from Monte-Carlo integration, the 3 dashed
curves are from the analytical theory.
The curves are labelled with
their value of $l/d$; the dashed curve labelled 0 is for spheres.
For all curves, $r_p=0.5r_c$ and $z=0.2$.
The inset graph is an enlargement of a small region of the main graph.
}
\label{fig3}
\end{figure}

\begin{figure}
\begin{center}
\epsfig{file=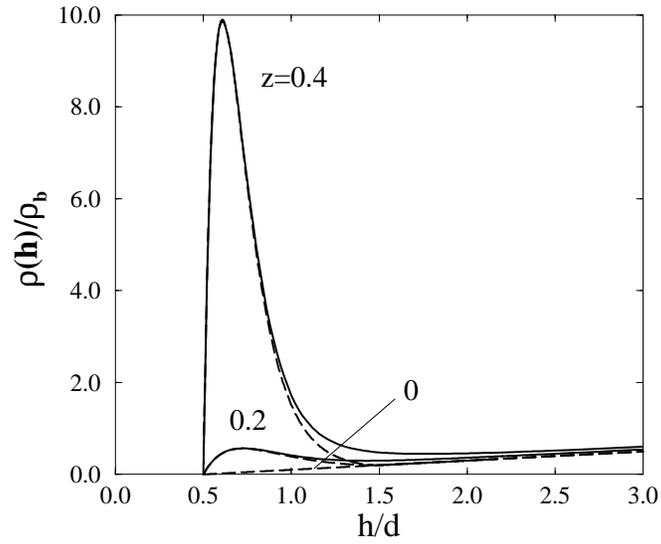, width=3.7in}
\end{center}
\vspace*{-0.7in}
\caption{
The ratio of the density of rods $\rho(h)$ to their bulk density $\rho_b$,
as a function of height $h$ above the wall.
The 2 solid curves are from Monte-Carlo integration, the 3 dashed
curves are from the analytical theory.
The curves are labelled with
their value of $z$.
For all curves, $r_p=r_c$ and $l/d=10$.
}
\label{fig4}
\end{figure}

\begin{figure}
\begin{center}
\epsfig{file=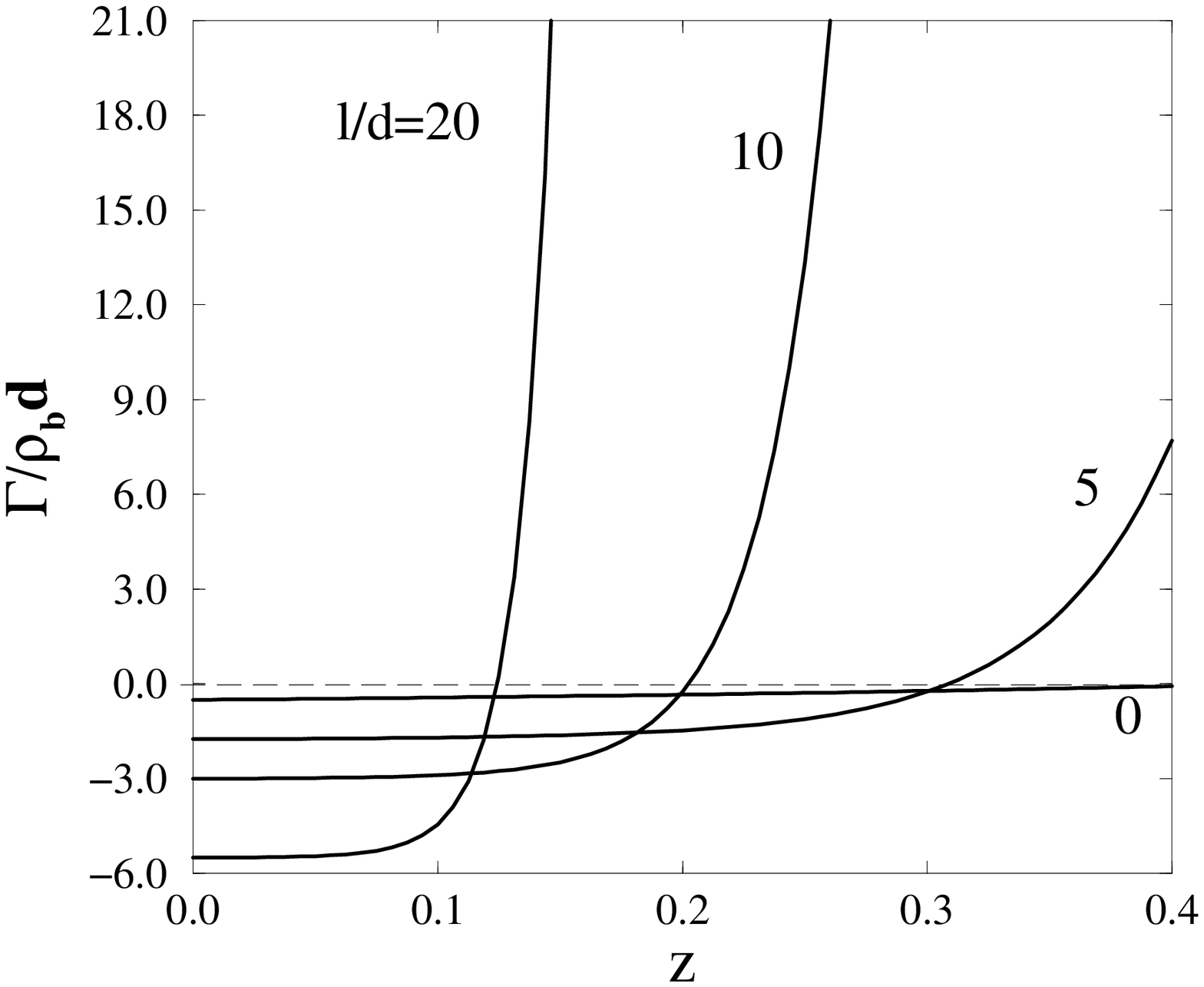, width=3.7in}
\end{center}
\vspace*{-0.7in}
\caption{
The adsorption $\Gamma$ is plotted as a function of $z$. Each curve
is labelled with its value of $l/d$. The curve labelled 0 is for spheres.
For all curves, $r_p=0.5r_c$.
}
\label{fig5}
\end{figure}

Now we are able to calculate the density of rods near a wall both
exactly via the numerical technique of Section IIB and via the simple
approximate theory of Section IIC.
We have done so for a number
of different values of the length to diameter ratio $l/d$ of the rods,
and the activity of the rods $z_p$. A dimensionless polymer
activity is required; we define the
reduced activity $z=z_p(2r_p)^3$.
As there are no
polymer--polymer interactions and the density of rods is so low that
they take up a negligible fraction of the volume, the density of polymer coils
is equal to their activity $z_p$ \cite{poon95,lekkerkerker95}.
So, the volume fraction occupied by the coils is
$\pi/6 z\simeq(1/2)z$.
In Figs. \ref{fig3} and \ref{fig4}, we show how the density
of the centres of mass of rods near
the wall depends on their length and on the concentration of the polymer.
Note that Fig. \ref{fig3} is for a polymer coils of radius half that
of the rod while Fig. \ref{fig4} is for polymer coils with the same
radius as the rod.
The free energy change on bringing
a rod from within the bulk to a height $h$ is proportional to
$-\ln\rho(h)$, see Eq. (\ref{freeh}).
The adsorption $\Gamma$ is plotted in Fig. \ref{fig5}.

Both the exact numerical integration of Eq. (\ref{rhoh}) over $\theta$ and
the approximate Eq. (\ref{rhoa}) are shown in Figs. \ref{fig3} and \ref{fig4}.
The approximation is seen to be excellent near the peak in $\rho(h)$. The
highest density of rods occurs only a little more than $r_c$ above the wall,
where the overlap of the excluded volumes is largest and the
orientations of a rod are very restricted.
This finding provides support for the related approximations made
in Ref. \cite{sear97}.
As can be seen in Fig. \ref{fig2} $v_o$ varies very little over the
restricted range of orientations available to a rod close to the wall.
This makes our approximation 
almost indistinguishable from the exact, numerical result at these heights.
Unlike spheres, the maximum in $\rho$ is not at $h=r_c$.
As $h$ decreases, the orientational freedom of the rod decreases,
see Eq. (\ref{rhoz0}).
and this tends to counterbalance the depletion attraction of the rod for the
wall. In the $h\rightarrow r_c$ limit the orientational freedom of a rod
tends to 0 and so the density of rods tends to 0.
The approximation is much less good near the minimum in $\rho(h)$, which
is to be expected as there the rods can rotate over a much larger angle.
The approximation consistently underestimates $\rho(h)$ because it
underestimates $v_o$. As can be seen in Fig. \ref{fig2}, $v_o$ is a minimum
for $\theta=90^{\circ}$.

From Fig. \ref{fig3} it is clear that, in the presence of polymer,
long rods, $l/d=10$,
adsorb much more strongly onto a hard wall than do spheres.
In Fig. \ref{fig3} the maximum density of spheres
is only about twice that in the bulk.
However, the density of rods reaches a maximum of over 20 times the bulk
value. For even longer rods the density maximum increases exponentially
and becomes very large. The larger density maxima correspond to free energies
of adhesion to the wall which are many times $k_BT$; rods much longer
than $l/d=10$ will adhere to the wall effectively irreversibly.
The sensitivity of the adsorption to the length of the rod is also
notable at small lengths. In Fig. \ref{fig3} we see that, for this value
of $z$, the density of short, $l/d=5$, rods is actually less than that of
spheres. As rods, but not spheres, approach a wall they lose orientational
entropy, which tends to reduce the density of the rods.
Thus, without polymer the density of rods but not of spheres is
reduced near a wall, see Eq. (\ref{rhoz0}).
For long rods and
high $z$ this is more than counterbalanced by the greater excluded volume
of rods as compared to spheres. As $z$ is increased beyond the value
of 0.2 the density of $l/d=5$ rods near the wall increases more rapidly
than that of spheres, and eventually becomes larger than that of spheres.

The dependence of the adsorption on the relative radii of the rod and of
the polymer may be seen if Figs. \ref{fig3} and \ref{fig4} are compared.
Fig. \ref{fig4} shows the density for rods with $l/d=10$ and polymer
of the same radius, $r_p=r_c$, while Fig. \ref{fig3} shows the density
of rods with the same $l/d$ and the same $z=0.2$ but with
polymer of radius half that of the rod, $r_p=0.5r_c$.
Of course, although the volume fraction of polymer is the same in
each case, the number density of the smaller polymer is 8 times that
of the larger.
If the polymer's radius is the same as that of the rod then for
$z=0.2$, which corresponds to a volume fraction of about 10\%, the
adsorption is very weak; the density near the wall is always less than that
in the bulk. However, the same volume fraction of polymer of half
the radius of the rods induces quite strong adsorption, a maximum density
over 20 times the bulk density.

The adsorption $\Gamma$, calculated using Eq. (\ref{gamma}), is plotted
in Fig. \ref{fig5}, as a function of $z$ for a number of values of $l/d$.
The curves were calculated using the approximate expression for $\rho(h)$,
Eq. (\ref{rhoa}).
For the calculation of the adsorption $\Gamma$ our approximation is
highly accurate as it is accurate where the density is highest.
This is particularly so at high $z$ or $l$,
the density in the peak near the wall grows exponentially with increasing
$z$ or $l$, see Eq. (\ref{rhoa}).
Then the adsorption is dominated by this peak.
It is obvious from Fig. \ref{fig5} that the adsorption of long rods is very
strong, or equivalently that only small volume fractions of polymer
are required to produce a layer of rods near the wall with a much higher
density than the bulk.

\section{Conclusion}

In a mixed suspension of rods and small polymer coils, the rods adsorb
onto a hard wall in contact with the suspension. The adsorption is
driven by depletion forces: The entropy of the polymer coils goes up when the
rods lie close to and parallel to the wall. From Figs. \ref{fig3} and
\ref{fig4} we see that the density of rods has a narrow peak, which implies
that a monolayer is formed. This is expected as the depletion forces are
short ranged.
The rod feels the wall up to $l/2+r_c+2r_p$ away but the interaction
is only strong when the rod approaches to within $r_c+2r_p$. Only then 
is there strong overlap of the excluded volumes. This is reflected
in the density profile. Just at the surface there is a dense monolayer
of rods lying parallel to the wall, then at larger separations the density
of rods is actually below that of the bulk.
The peak in the density is strongly dependent on $l$; longer rods
adsorb strongly onto the wall. Again this is unsurprising, the
excluded volume and hence the potential increase in the entropy of the
polymer coils increases linearly with the length of the rod.

If we compare rods with spheres, we see that in the absence of polymer
the density of rods but not of spheres drops as the wall is approached.
This is due to the decrease in orientational freedom as a rod approaches
a wall. In the presence of polymer, the larger excluded volume of a
rod as compared with a sphere (if they have the same radius)
causes the density of rods near the wall to increase more quickly than
that of rods. The density of long rods becomes much higher near the wall
than does the density of spheres.

Equation (\ref{rho}) for the density has exactly the same form for
a rod attracted to the wall by a direct rod--wall interaction.
So, in the presence of any attraction with a range
approximately equal to the diameter of the rod, the density profiles
will be qualitatively the same as those of our Figs. \ref{fig3} and \ref{fig4}.
This applies to, for example,
van der Waals attractions \cite{schoot92,sear97}.

So far, we have only considered an ideal gas of rods adsorbing onto
a wall. This was reduced to a simple one-particle problem which, of course,
did not show a phase transition.
Although we will not perform any calculations which account for interactions
between rods at the wall it is, perhaps, interesting to speculate
on what will happen near the wall.
We have seen that the density
near the wall can become much higher than that in the bulk, particularly
if $l/d$ is large. The density peak is narrow due to the short ranged
attraction and so only a monolayer is formed. So, we expect a dense
monolayer, which would resemble a system of 2-dimensional rods.
For strong adsorption the density of this monolayer will be
much higher than in the bulk.
This immediately suggests that even at low bulk densities,
i.e., densities below a bulk transition to a nematic
phase \cite{vroege92} or between
two fluid phases \cite{lekkerkerker94,bolhuis97},
the density of a monolayer just above the wall may be high enough
for this monolayer to undergo a transition to a 2-dimensional
nematic phase \cite{frenkel91,matsuyama97,wang90}.
We expect the surface layer to be in the nematic phase for densities
$\Gamma$ of order $1/ld$ and greater.
Crystallisation of spheres at a wall at lower bulk densities than
required for bulk crystallisation have been observed \cite{kaplan94} and
treated theoretically \cite{poon94}.

\section*{Acknowledgements}

It is a pleasure to acknowledge
a careful reading of the manuscript by J. Doye.
I would like to thank The Royal Society for the award of a fellowship
and the FOM institute AMOLF for its hospitality.
The work of the FOM Institute is part of the research program of FOM
and is made possible by financial support from the
Netherlands Organisation for Scientific Research (NWO).


\end{multicols}
\widetext

\newpage

\newpage

\newpage

\newpage

\newpage


\begin{thebibliography}{99}

\bibitem{schoot92} P. van der Schoot and T. Odijk,
J. Chem. Phys. {\bf 97} (1992) 515

\bibitem{sear97} R. P. Sear, Phys. Rev. E {\bf 55}, 5820 (1997).

\bibitem{frenkel91} D. Frenkel, in
{\it Liquids, Freezing and Glass Transitions, Les Houches session},
edited by J.-P. Hansen, D. Levesque and J. Zinn-Justin
(North-Holland, Amsterdam, 1991).

\bibitem{allen94} M. P. Allen, G. T. Evans, D. Frenkel and B. M. Mulder,
Adv. Chem. Phys. {\bf 86}, 1 (1994).

\bibitem{gast83} A. P. Gast, C. K. Hall and W. B. Russell,
{\it Faraday Discuss. Chem. Soc.} {\bf 76}, 189 (1983).

\bibitem{lekkerkerker92} H. N. W. Lekkerkerker, W. C. K. Poon, P. N. Pusey,
A. Stroobants and P. B. Warren, Europhys. Lett. {\bf 20}, 559 (1992).

\bibitem{russel89} W. B. Russel, D. A. Saville and W. R. Schowalter,
{\it Colloidal Dispersions} (Cambridge University Press,
Cambridge, 1989).

\bibitem{poon95} W. C. K. Poon and P. N. Pusey, in
{\it Proceedings of the International School of Physics \lq Enrico Fermi'
Course CXXIX}, edited by M. Baus, L. F. Rull and J. P. Ryckaert
(Kluwer, Dordrecht, 1995).

\bibitem{lekkerkerker95} H. N. W. Lekkerkerker, P. Buining, J. Buitenhuis,
G. J. Vroege and A. Stroobants,
{\it Proceedings of the International School of Physics \lq Enrico Fermi'
Course CXXIX}, edited by M. Baus, L. F. Rull and J. P. Ryckaert
(Kluwer, Dordrecht, 1995).

\bibitem{kaplan94b} P. D. Kaplan, L. P. Faucheux and A. J. Libchaber, 
Phys. Rev. Lett. {\bf 73}, 2793 (1994).

\bibitem{sober95} D. L. Sober and J. Y. Walz,
Langmuir {\bf 11}, 2352 (1995).

\bibitem{dinsmore96} A. D. Dinsmore, A. G. Yodh and D. J. Pine,
Nature {\bf 383}, 239 (1996).

\bibitem{kaplan94} P. D. Kaplan, J. L. Rourke, A. G. Yodh and
D. J. Pine, Phys. Rev. Lett. {\bf 72}, 582 (1994).

\bibitem{ohshima97} Y. N. Ohsima {\it et al},
Phys. Rev. Lett. {\bf 78}, 3963 (1997).

\bibitem{asakura54} S. Asakura and F. Oosawa,
J. Chem. Phys. {\bf 22},
1255 (1954).

\bibitem{vrij76} A. Vrij,
Pure Appl. Chem. {\bf 48}, 471 (1976).

\bibitem{eisenriegler96} E. Eisenriegler, A. Hanke and S. Dietrich,
Phys. Rev. E {\bf 54}, 1134 (1996).

\bibitem{fraden95} S. Fraden,
{\it Proceedings of the International School of Physics \lq Enrico Fermi'
Course CXXIX}, edited by M. Baus, L. F. Rull and J. P. Ryckaert
(Kluwer, Dordrecht, 1995).

\bibitem{buining94} P. A. Buining, A. P. Philipse and H. N. W. Lekkerkerker,
Langmuir {\bf 10}, 2106 (1994).

\bibitem{buitenhuis95} J. Buitenhuis, L. N. Donselaar, P. A. Buining,
A. Stroobants and H. N. W.  Lekkerkerker,
J. Colloid Interface Sci. {\bf 175}, 46 (1995).

\bibitem{matsuyama97} A. Matsuyama, R. Kishimoto and T. Kato,
J. Chem. Phys. {\bf 106}, 6744 (1997).

\bibitem{linden96} C. C. van der Linden, F. A. M. Leermakers and
G. J. Fleer, Macromolecules {\bf 29}, 1172 (1996).

\bibitem{vroege92} G. J. Vroege and  H. N. W. Lekkerkerker,
Rep. Prog. Phys. {\bf 55}, 1241 (1992).

\bibitem{warren94} P. B. Warren,
J. Phys. France I {\bf 4}, 237 (1994) 237.

\bibitem{lekkerkerker94} H. N. W. Lekkerkerker and A. Stroobants,
Nuovo Cimento {\bf 16D}, 949 (1994).

\bibitem{rowlinson82} J. R. Rowlinson and B. Widom,
{\it Molecular Theory of Capillarity} (Clarendon Press, Oxford, 1982).

\bibitem{press92} W. H. Press, S. A. Teukolsky, W. T. Vetterling, and
B. P. Flannery,
{\it Numerical Recipes} (Cambridge University Press, Cambridge,
2nd ed, 1992).

\bibitem{allen87} M. P. Allen and D. J. Tildesley,
{\it Computer Simulations of Liquids} (Clarendon Press,
Oxford, 1987).

\bibitem{sear97b} R. P. Sear, J. Phys. France II {\bf 7}, 877 (1997).

\bibitem{bolhuis97} P. G. Bolhuis, A. Stroobants, D. Frenkel and
H. N. W. Lekkerkerker, J. Chem. Phys. {\bf 107}, 1551 (1997).

\bibitem{wang90} Z. G. Wang, J. Phys France {\bf 51}, 1431 (1990).

\bibitem{poon94} W. C. K. Poon and P. B. Warren,
Europhys. Lett. {\bf 28}, 513 (1994)




\end{thebibliography}
\end{document}